\documentclass{JHEP3}

\usepackage{Pformel,Pmeta}
\usepackage{amsmath,amsfonts,amsthm,times}

\title{The large central charge limit of conformal blocks}

\author{Vladimir Fateev$^{1,2}$ and Sylvain Ribault$^1$
\\
\!\!\!$^1$\! Laboratoire Charles Coulomb UMR 5221 CNRS-UM2
\\
Universit\'e Montpellier 2, Place Eug\`ene Bataillon - CC070
\\
F-34095 Montpellier Cedex 5 - France
 \\
\!\!\!$^2$\! Landau Institute for Theoretical Physics
 \\
 142432 Chernogolovka, Russia
 \\
 {\footnotesize \tt vladimir.fateev@univ-montp2.fr, sylvain.ribault@univ-montp2.fr }
}

\abstract{We study conformal blocks of conformal field theories with a $W_3$ symmetry algebra in the limit where the central charge is large. In this limit, we compute the four-point block as a special case of an $s\ell_3$-invariant function. In the case when two of the four fields are semi-degenerate, we check that our results agree with the block's combinatorial expansion as a sum over Young diagrams. We also show that such a block obeys a sixth-order differential equation, and that it has an unexpected singularity at $z=-1$, in addition to the expected singularities at $z=0,1,\infty$.
}

\makeatletter\let\default@color\current@color\makeatother 

\preprint{}

\begin{document}

\zeq\section{Introduction}

Since the work of Belavin, Polyakov and Zamolodchikov \cite{bpz84}, the conformal bootstrap method has been an effective tool for studying two-dimensional conformal field theories. That method relies on a systematic exploitation of the symmetries of the theory. These symmetries determine functions called conformal blocks. The simplest conformal blocks are the characters of the representations of the symmetry algebra, which may be called zero-point blocks on the torus. The correlation functions of the theory are then combinations of the conformal blocks. For example, the partition function on the torus is a combination of characters. 

Combinatorial expansions for the conformal blocks of the Virasoro algebra have recently been found \cite{aflt10}, inspired by the conjecture of Alday, Gaiotto and Tachikawa on the relation between two-dimensional CFTs and four-dimensional gauge theories \cite{agt09}. Until then, no explicit formulas for four-point blocks on the sphere were known. It would be very interesting to generalize such combinatorial expansions to conformal blocks of other algebras, in particular the $W_N$ algebras \cite{zam85,fz86b,fl88} which are natural generalizations of the Virasoro algebra.

This is however a challenging problem, in particular because the fusion products of $W_{N\geq 3}$ representations in general exhibit infinite fusion multiplicites. (See for instance Section 2.3 of \cite{fr10}.) This feature is at the origin of difficulties in computing the three-point correlation functions in conformal Toda theories. These CFTs have $W_N$ symmetry algebras, and their three-point correlation functions are only known in special cases \cite{fl07c,fl08}. Here we will show how to take infinite fusion multiplicities into account and how to compute conformal blocks in the limit where 
the central charge $c$ of the $W_N$ algebra is large. (The conformal dimensions of the fields are meanwhile kept fixed; this is sometimes called the light asymptotic limit.) It is this limit which was used by Al. Zamolodchikov as the starting point of the characterization of Virasoro ($N=2$) conformal blocks by recurrence \cite{zam84}. 
In this limit, $s\ell_N$ conformal Toda theory reduces to the quantum mechanics of a point particle on $SL_N(\C)$, the $W_N$ algebra reduces to $s\ell_N$, and the $W_N$ conformal blocks reduce to special cases of $s\ell_N$-invariant functions. We will study such functions in detail in the cases $N=2$ and $N=3$. 

This will enable us to test a proposal for the combinatorial expansion of a class of $W_N$ conformal blocks \cite{fl11}. The proposed expansion, which we will summarize, is given for all values of $c$, assuming that all involved fields except two of them are almost fully degenerate. This assumption eliminates the problem of the infinite fusion multiplicities. We will compare our results for large $c$ conformal blocks of that class, with the large $c$ limit of the combinatorial expansion. The two expansions agree up to the order $z^5$, which supports the validity of the proposed combinatorial expansion.

\paragraph{Plan of the article. } In Section \ref{smain}, after a reminder on the large $c$ limit of Virasoro conformal blocks, we study $s\ell_3$-invariant functions and their relations with the large $c$ limit of $W_3$ conformal blocks. 
Section \ref{appdc} is devoted to the study of detailed properties of certain conformal blocks: the differential equation they obey, and their critical exponents.
Then, in the concluding Section \ref{sconc}, we comment on some aspects of the results, and compare them with the combinatorial expansion. Appendix \ref{appqm} is devoted to the study of a quantum particle on $SL_N(\C)$ (with $N=2,3$), which is at the basis of the computation of the large $c$ limit of correlation functions of $s\ell_N$ conformal Toda theory, and provides some justification for a number of the equations of Section \ref{smain}. Appendix \ref{appder} is devoted to deriving the series expansion (\ref{fsss}) of certain  conformal blocks. 

\paragraph{Acknowledgements.} We thank Vladimir Belavin for comments on the draft of this article. We are grateful to the JHEP referee for suggestions which led to significant improvements. This work was supported in part by the cooperative CNRS-RFBR grant PICS-09-02-93106. S. R. is grateful to the Institut Poncelet in Moscow for hospitality while part of this work was done.

\section{$W_3$ conformal blocks and $s\ell_3$-invariant functions \label{smain}}

\subsection{Virasoro conformal blocks and $s\ell_2$-invariant functions}

We first review the case of Virasoro conformal blocks, before moving to the technically more complicated case of $W_3$ conformal blocks. Basic information on Virasoro conformal blocks can be found in \cite{fms97}. 
A four-point $s$-channel Virasoro conformal block on the sphere ${\cal G}_{\Delta_s}(c|\Delta_i|z_i)$ is a function of the positions $(z_1,z_2,z_3,z_4)\in \C^4$ of four primary fields, which are characterized by their conformal dimensions $(\Delta_1,\Delta_2,\Delta_3,\Delta_4)$. Such a conformal block also depends on an $s$-channel conformal dimension $\Delta_s$, and on the central charge $c$ of the Virasoro algebra, which is defined by generators $L_{n\in \Z}$ and relations
\bea
[L_n,L_m] = (n-m)L_{m+n} + \frac{c}{12} n(n-1)(n+1) \delta_{m+n,0} \ .
\eea
The conformal block is defined as a sum over the states of a highest-weight representation of the Virasoro algebra. 
The relevant representation is built from a highest-weight state $|\Delta_s\rangle$ 
by applying the creation modes $L_{n<0}$. If we assume $L_n^\dagger = L_{-n}$ and $\langle \Delta_s| \Delta_s \rangle = 1$, then we can compute the square norm $|| L_{-n} |\Delta_s\rangle ||^2 = 2n\Delta_s + \frac{c}{12} n(n-1)(n+1)$. If $|n|\geq 2$, this goes to infinity as $c\rar \infty$, and $L_{-n}$ descendents 
do not contribute to the "large $c$ block"
\bea
{\cal F}_{\Delta_s}(\Delta_i|z_i) = \underset{c\rar \infty}{\lim} {\cal G}_{\Delta_s}(c|\Delta_i|z_i) \ .
\label{flg}
\eea
Therefore, only the generators $(L_{-1},L_0,L_1)$ of the  $s\ell_2$ subalgebra of global conformal transformations survive in the large $c$ limit. Nonetheless, some properties of the blocks are not affected by taking this limit: 
First, the existence of an analytic expansion in the neighbourhood of $z_1=z_2$,
\bea
{\cal F}_{\Delta_s}(\Delta_i|z_i) =
z_{12}^{\Delta_s-\Delta_1-\Delta_2}(1+O(z_{12}))\ ,
\label{fzo}
\eea
where we use the notation $z_{12}=z_1-z_2$. Second, the behaviour under global conformal transformations, which we now review.

The $s\ell_2$ subalgebra of global conformal transformations has the
  generators $(L_{-1},L_0,L_1)$ and commutation relations
\bea
[L_0,L_{\pm 1}] = \mp L_{\pm 1} \scs  [L_1,L_{-1}] = 2L_0\ .
\label{lll}
\eea
A representation of the Virasoro algebra with conformal dimension $\Delta$ corresponds to an $s\ell_2$ representation of spin $-\Delta$.
Notice that two Virasoro representations whose dimensions are related by the reflection
\bea
\Delta^* = 1-\Delta \ ,
\label{dod}
\eea
correspond to two isomorphic $s\ell_2$ representations. 
A primary field with position $z$ and conformal dimension $\Delta$ behaves as a vector in an $s\ell_2$ representation of spin $-\Delta$ and isospin variable $z$. 
The action of the $s\ell_2$ subalgebra on the primary field is given by the differential operators
\bea
D_{(\Delta,z)}(L_{-1}) = -\pp{z} \scs D_{(\Delta,z)}(L_0) = -z\pp{z} - \Delta
 \scs D_{(\Delta,z)}(L_{1}) = - z^2 \pp{z} - 2\Delta z \ ,
\label{ddd}
\eea
which are such that $D_{(\Delta,z)}$ preserves the commutation relations (\ref{lll}). 
The blocks ${\cal G}_{\Delta_s}(c|\Delta_i|z_i)$, and therefore their large $c$ limits ${\cal F}_{\Delta_s}(\Delta_i|z_i)$, are $s\ell_2$-invariant four-points functions. What we call an $s\ell_2$-invariant $n$-points function is a function ${\cal E}(\Delta_i|z_i)$ of $(\Delta_1,\Delta_2,\cdots \Delta_n)$ and $(z_1,z_2,\cdots z_n)$ such that
\bea
\forall\ t^a \in \{L_{-1},L_0,L_1\}, \qquad
\left(\sum_{i=1}^n D_{(\Delta_i,z_i)}(t^a) \right) {\cal E}(\Delta_i|z_i) = 0 \ .
\label{tlll}
\eea
The invariant two- and three-points functions are well-known to be
\bea
{\cal E}(\Delta_1,\Delta_2|z_1,z_2) &=& z_{12}^{-2\Delta_1} \scs ({\rm assuming } \ \Delta_1=\Delta_2)\ ,
\label{edd}
\\
{\cal E}(\Delta_1,\Delta_2,\Delta_3|z_1,z_2,z_3) &=&
z_{12}^{\Delta_{3}-\Delta_{1}-\Delta_{2}} z_{23}^{\Delta_{1}-\Delta_{2}-\Delta_{3}} z_{31}^{\Delta_{2}-\Delta_{3}-\Delta_{1}} \ .
\label{eddd}
\eea
Any invariant four-points function ${\cal E}(\Delta_i|z_i)$ can be written in terms of its 
values when three of the $z_i$s are fixed, for example $(z_1,z_3,z_4)=(0,1,\infty)$,
\bea
{\cal E}(\Delta_i|z_i) = P(\Delta_i|z_i) z^{\Delta_1+\Delta_2}{\cal E}(\Delta_i|0,z,1,\infty) \ , 
\label{fog}
\eea
where we define the cross-ratio $z$  and prefactor $P(\Delta_i|z_i)$ as
\bea
P(\Delta_i|z_i)=
z_{12}^{-\Delta_1-\Delta_2} z_{13}^{-\Delta_3+\Delta_4}
z_{43}^{-\Delta_3-\Delta_4}z_{42}^{\Delta_1-\Delta_2}
z_{14}^{-\Delta_1+\Delta_2+\Delta_3-\Delta_4} 
\scs 
z = \frac{z_{12}z_{34}}{z_{13}z_{24}} \ .
\label{oz}
\eea
(Notice that we have $P(\Delta_i|z_i) = {\cal E}(\Delta_1,\Delta_2,0|z_1,z_2,z_4){\cal E}(0,\Delta_3,\Delta_4|z_1,z_3,z_4)$.) For brevity we will sometimes use the notation ${\cal E}(z) = {\cal E}(\Delta_i|0,z,1,\infty)$.

After these reminders on the global conformal symmetry, we are ready to write an explicit integral formula for the large $c$ four-point conformal block,
\bea
{\cal F}_{\Delta_s}(\Delta_i|z_i) = {\cal N} \int_C dz_s\ {\cal E}(\Delta_1,\Delta_2,\Delta_s|z_1,z_2,z_s) {\cal E}(\Delta_s^*,\Delta_3,\Delta_4|z_s,z_3,z_4) \ ,
\label{fds}
\eea
where the normalization factor ${\cal N}$ (a function of $\Delta_i,\Delta_s$) and integration contour $C$ are determined by the condition (\ref{fzo}). This expression is justified in Appendix \ref{saqm}. With a general integration contour, the integral in eq. (\ref{fds}) would yield a linear combination of the two "reflected" blocks ${\cal F}_{\Delta_s}$ and ${\cal F}_{\Delta_s^*}$.
Explicitly, we find
\bea
{\cal F}_{\Delta_s}(\Delta_i|0,z,1,\infty) &= &
z^{-\Delta_{12}^s} \sum_{n=0}^\infty \frac{(\Delta_s-\Delta_1+\Delta_2)_n(\Delta_s-\Delta_4+\Delta_3)_n}{(2\Delta_s)_n} z^n \ ,
\\
&=&
z^{-\Delta_{12}^s} F(\Delta_s-\Delta_1+\Delta_2,\Delta_s-\Delta_4+\Delta_3,2\Delta_s,z) \ ,
\label{zf}
\eea
where $F$ is the hypergeometric function, and we use the notations
\bea
(t)_n =\prod_{i=0}^{n-1} (t+i) =\frac{\G(t+i)}{\G(t)} \ ,
\label{tn}
\eea
and 
\bea
\Delta_{12}^s = \Delta_1+\Delta_2-\Delta_s\ .
\eea

The simplification of the conformal blocks in the large $c$ limit can be interpreted as coming from the elimination of local conformal symmetry, and the survival of only the global symmetry with its finite-dimensional $s\ell_2$ algebra. This explains why
the large $c$ blocks (\ref{zf}) coincide with the conformal partial waves which were computed by Ferrara, Gatto and Grillo \cite{fgg75}. 
Such conformal partial waves are associated to the global conformal symmetry, and can therefore be generalised to higher dimensions \cite{do93}\footnote{We are grateful to Slava Rychkov for pointing out the articles \cite{fgg75} and \cite{do93} to us.}. We will be interested in another type of generalisation: staying in two dimensions, we will consider larger symmetry algebras.

\subsection{$s\ell_3$-invariant functions \label{ssif}}

In preparation for writing the large $c$ conformal blocks of the $W_3$ algebra, we need to study $s\ell_3$-invariant functions. This is because the $W_3$ algebra reduces to $s\ell_3$ in the large $c$ limit \cite{bw91}\footnote{According to \cite{bw91} (pages 7-8), $W_3$ reduces to $s\ell_3$ by a two-step process of truncating to the vacuum-preserving algebra and taking the large $c$ limit. However, for our purpose of computing large $c$ conformal blocks, the large $c$ limit does perform the truncation, as we explained in the case of the Virasoro algebra. }, in the same way as the Virasoro algebra reduces to $s\ell_2$. 
 (The eight generators $L_0,L_{\pm 1},W_0,W_{\pm 1},W_{\pm 2}$ of the $W_3$ algebra which survive in the large $c$ limit can be identified with linear combinations of the generators $h^i,e^i,f^i$ of $s\ell_3$ which we are about to introduce.)

The algebra $s\ell_3$ is generated by two Cartan elements $(h^1,h^2)$, three generators $(f^1,f^2,f^3)$ which are eigenvectors of the adjoint actions of $h^1$ and $h^2$ for the respective eigenvalues $(-2,1,-1)$ and $(1,-2,-1)$, and three generators $(e^1,e^2,e^3)$ which are also eigenvectors but with opposite eigenvalues. The remaining nonzero commutators are 
\bea 
&&
[f^1,f^2]=-f^3 \scs  [e^1,e^2]=e^3\ ,
\\
&&
[e^1,f^1]=h^1 \scs [e^2,f^2]=h^2 \scs   [e^3,f^3]=h^1+h^2 \ ,
\\
&&
[e^1,f^3]=-f^2 \scs [e^2,f^3]=f^1 \scs [e^3,f^1]=-e^2 \scs [e^3,f^2]=e^1\ .
\label{hef}
\eea
In order to parametrize the representations of $s\ell_3$, let us introduce its simple roots $(e_1,e_2)$ and the weights of the fundamental representation $(h_1,h_2,h_3)$ (not to be confused with the $s\ell_3$ generators $e^i,h^j$). The roots are supposed to be two independent vectors, with a scalar product given by the Cartan matrix, $(e_i,e_j)=K_{ij}$ with $K=\bsm 2 & -1 \\ -1 & 2 \esm$. The weights of the fundamental representation are 
\bea
h_1= \tfrac23 e_1 +\tfrac13 e_2 \scs h_2= -\tfrac13 e_1 +\tfrac13 e_2 \scs h_3 = -\tfrac13 e_1 -\tfrac23 e_2 \ .
\label{hhh}
\eea
A representation is parametrized by a spin vector $j$ in root space, whose coordinates we denote as
\bea
r = -(e_1,j) \scs s = -(e_2,j) \ .
\label{rjsj}
\eea
Two representations are isomorphic when they are related by one of the six Weyl transformations
\bea
(r,s) \ \rar \  \left\{
\begin{array}{lll}
 (r,s), & (3-r-s,r), & (s,3-r-s), 
\\
 (2-r,-1+r+s), &(-1+r+s,2-s),  &(2-s,2-r), 
\end{array}
\right. 
\label{rsr}
\eea
among which we single out the maximal Weyl reflection $j\rar j^*$ where
\bea
j = (r,s) \Rightarrow j^* = (2-s,2-r)\ .
\label{jjs}
\eea
There is another useful reflection of the root space called the Dynkin diagram automorphism $j\rar j^\omega$ where
\bea
j= (r,s) \Rightarrow j^\omega =  (s,r) \ .
\label{jje}
\eea

In the previous Subsection, we represented $s\ell_2$ transformations in terms of differential operators (\ref{ddd}), whose isospin variable $z$ could be interpreted as the position of a CFT field on the complex plane. In order to faithfully represent $s\ell_3$ transformations, we need a triple of variables $Z=(w,x,y)$. (The number of needed variables is the number of creation operators $e^i$; in the case of $s\ell_N$ this would be $\frac{N(N-1)}{2}$.) 
The $s\ell_3$ generators $(h^i,e^i,f^i)$ are represented as \cite{fms97}(Section 15.7.4)
\bea
D_{(j,Z)}(h^1)& =& 2 x \p_x+r -y \p_y + w \p_w \ ,
\label{dh}
\\
D_{(j,Z)}(h^2) &=& 2 y \p_y+s -x \p_x + w \p_w \ ,
\\
D_{(j,Z)}(e^1)& =&  x^2 \p_x+r x + (w-x y) \p_y + x w \p_w \ ,
\\
D_{(j,Z)}(e^2) &=&  y^2 \p_y+s y -w \p_x \ ,
\\
D_{(j,Z)}(e^3)& =&  w^2 \p_w+s (w-x y)+r w+ x w \p_x 
 +y (w-x y) \p_y \ ,
\\
D_{(j,Z)}(f^1)& =& - \p_x \ ,
\\
D_{(j,Z)}(f^2)&= &- \p_y -x \p_w\ ,
\\
D_{(j,Z)}(f^3) &=& - \p_w \ .
\label{df}
\eea
An $s\ell_3$-invariant $n$-point function associated to $n$ spins $j_1,j_2,\cdots j_n$ is a function ${\cal E}(j_i|Z_i)$ such that
\bea
\forall\ t^a\in \{h^i,e^i,f^i\}, \quad \left(\sum_{i=1}^n D_{(j_i,Z_i)}(t^a)\right){\cal E}(j_i|Z_i) = 0\ .
\label{sde}
\eea
Such an invariant will obey additional equations if some representations are degenerate. We will call the representation of spin $j_1$ semi-degenerate of the first $(k=1)$ or second $(k=2)$ kind if 
\bea
(e_k,j_1)= 0 \qquad {\rm and} \qquad d^{(k)}_{Z_1} {\cal E}(j_i|Z_i) = 0 \ ,
\label{ded}
\eea
where the differential operators $d^{(k)}_Z$ are defined as 
\bea
d^{(1)}_Z = \p_x +y\p_w \scs d^{(2)}_Z  = \p_y \ .
\label{dzdz}
\eea 
These formulas for $d^{(1)}_Z$ and $d^{(2)}_Z$ will be justified in Appendix \ref{sapp}.

Let us write the solutions of the $s\ell_3$ invariance equation (\ref{sde}) in the cases of two- and three-point invariants. We will write the solutions of these equations in terms of convenient combinations of isospin variables $Z_i=(w_i,x_i,y_i)$,
\bea
\rho_{ij} &=&  y_i(x_i-x_j) - (w_i-w_j) \ , 
\label{rho}
\\
\sigma_{ijk} & = & x_iw_j-x_jw_i+x_jw_k-x_kw_j+x_kw_i-x_iw_k \ ,
\\
\chi_{ijk} & = & y_iw_j-y_jw_i+y_jw_k-y_kw_j+y_kw_i-y_iw_k 
\nn
\\ & & 
+y_iy_j(x_i-x_j) +y_jy_k(x_j-x_k) + y_ky_i(x_k-x_i) \ .
\label{chi}
\eea
We also introduce the three-point invariant
\bea
\theta_{ijk} = \frac{\rho_{ij}\rho_{jk}\rho_{ki}}{\rho_{ji}\rho_{kj}\rho_{ik}} \quad \Rightarrow \quad 
 \left(D_{(0,Z_i)}(t^a)+ D_{(0,Z_j)}(t^a)+D_{(0,Z_k)}(t^a)\right) \theta_{ijk} = 0 \ .
\label{tst}
\eea
Our combinations are related by identities of the type
\bea
&& \sigma_{123}\chi_{123} = \rho_{21}\rho_{32}\rho_{13} \left(\theta_{123}+1\right) \ ,
\\
&& \chi_{123} \rho_{41} + \chi_{134}\rho_{21} + \chi_{142}\rho_{31} = 0\ .
\eea
We then find that a nonzero two-point invariant can exist only provided $j_1 = j_2^\omega$ (up to Weyl reflections), and the invariant is then
\bea
{\cal E}(j_1,j_2|Z_1,Z_2) = \rho_{21}^{-r_1} \rho_{12}^{-s_1} \scs ({\rm assuming } \ j_1=j_2^\om) \ .
\label{fj}
\eea
Consider now three-point invariants. The function ${\cal E}(j_1,j_2,j_3|Z_1,Z_2,Z_3)$ depends on nine variables which are the components of $Z_1,Z_2,Z_3$, and is subject to the eight equations (\ref{sde}). Therefore, there exists an infinite-dimensional space of solutions. This corresponds to the existence of a nontrivial invariant $\theta_{123}$ (\ref{tst}). In the special case when one of the three representations is semi-degenerate, we have an extra equation of the type (\ref{ded}), and the space of solutions is one-dimensional. Let us start with the case when the first representation is semi-degenerate of the first kind. The three-point invariant should be built from  combinations of isospin variables which are killed by the differential operator $d^{(1)}_{Z_1}$ (\ref{dzdz}), for instance
\bea
d^{(1)}_{Z_1} \rho_{12} = 0 \scs d^{(1)}_{Z_1} \chi_{123} = 0 \ .
\eea
We then find the three-point invariant
\bea
{\cal E}(j_1,j_2,j_3|Z_1,Z_2,Z_3) = \chi_{123}^{-J} \rho_{12}^{-J-r_2+s_3} \rho_{13}^{-J-r_3+s_2} \rho_{23}^{J-s_2} \rho_{32}^{J-s_3} \ , \qquad (r_1=0) \ , 
\label{frz}
\eea
where we introduced the combination of spins
\bea
J = (h_2,j_1+j_2+j_3) = \tfrac13(s_1+s_2+s_3-r_1-r_2-r_3)\ .
\label{jhj}
\eea
Similarly, if the first representation is semi-degenerate of the second kind, we can use combinations which are killed by $d^{(2)}_{Z_1}$, in particular
\bea
d^{(2)}_{Z_1} \rho_{21} = 0 \scs d^{(2)}_{Z_1} \sigma_{123} = 0 \ ,
\eea
and we find the three-point invariant
\bea
{\cal E}(j_1,j_2,j_3|Z_1,Z_2,Z_3) = \sigma_{123}^{J} \rho_{21}^{J+r_3-s_2} \rho_{31}^{J+r_2-s_3} \rho_{23}^{-J-r_3} \rho_{32}^{-J-r_2}\ , \qquad (s_1=0)\ .
\label{fsz}
\eea
In general, when no representation is semi-degenerate, the most general three-point invariant is
\bea
{\cal E}_{g_1}(j_i|Z_i) &=&  \chi_{123}^{-J} \rho_{12}^{-J-r_1-r_2+s_3} \rho_{13}^{-J-r_3+s_2} \rho_{23}^{J-s_2} \rho_{32}^{J+r_1-s_3} \rho_{31}^{-r_1}\ g_1(\theta_{123})\ ,
\label{egc}
\\
 &=& \sigma_{123}^{J} \rho_{21}^{J+r_3-s_1-s_2} \rho_{31}^{J+r_2-s_3} \rho_{23}^{-J-r_3+s_1} \rho_{32}^{-J-r_2}\rho_{13}^{-s_1}\ g_2(\theta_{123})\ .
\eea
This depends on an arbitrary ``multiplicity function'' $g_1(\theta)$, or on the equivalent function  $g_2(\theta)=\theta^{-J-r_1-r_2+s_3} {} (\theta+1)^{-J}g_1(\theta)$. This function encodes the infinite multiplicity of say the third representation in the tensor product of the first two representations. The same feature manifests itself in the fusion products of $W_3$ representations, we have called this the problem of the infinite fusion multiplicities in the Introduction.

Bases of $n$-point invariants can be built from three-point invariants. For instance, $s$-channel four-point invariants can be built as 
\bea
{\cal E}_{g,g'|j_s}(j_i|Z_i) = {\cal N}\int_C dZ_s\ {\cal E}_g(j_1,j_2,j_s|Z_1,Z_2,Z_s) {\cal E}_{g'}(j_s^{*\om},j_3,j_4|Z_s,Z_3,Z_4)\ ,
\label{egg}
\eea
where $j_s$ is the $s$-channel spin, $g,g'$ are two multiplicity functions, ${\cal N}$ is a normalization factor which may depend on $j_s,j_i,g,g'$, and $C$ is an integration domain for $Z_s\in \C^3$. The integration measure is the $s\ell_3$-invariant measure $dZ = dw dx dy$.

\subsection{$W_3$ conformal blocks in the large $c$ limit \label{sscbl}}

A four-point $s$-channel $W_3$ conformal block on the sphere ${\cal G}_{g,g'|\al_s}(c|\al_i|z_i)$ is a function of the positions $(z_1,z_2,z_3,z_4)$ of four primary fields characterized by their momenta $(\al_1,\al_2,\al_3,\al_4)$, and of the central charge $c$ of the $W_3$ algebra. The block also depends on an $s$-channel momentum $\al_s$, and on two multiplicity functions $g,g'$. The presence of such multiplicity functions is in general necessary due to the presence of infinite fusion multiplicities, and we have given a precise definition of such multiplicity functions in the case of $s\ell_3$-invariant functions in the previous Subsection. We will however not try to define such functions in the case of $W_3$ conformal blocks, except in the large $c$ limit. 

Let us introduce standard notations on $W_3$ representations. Let $b$ and $q=b+b^{-1}$ be such that the central charge is $c=2+24q^2$. 
A highest-weight representation of the $W_3$ algebra is parametrized by its momentum $\al$, a two-dimensional vector which belongs to the root space of $s\ell_3$. Such a representation can alternatively be parametrized by its conformal dimension $\Delta_\al$ and a charge $q^{(3)}_\al$ which is (up to a normalization factor) the eigenvalue of the spin $3$ current, such that
\bea
\Delta_\al = \frac12(\al,2Q-\al) \scs q^{(3)}_\al = -3b\prod_{i=1}^3(h_i,\al-Q)\ , 
\eea
where we defined $Q=q\rho$, which involves the Weyl vector $\rho = e_1+e_2$. 

The large $c$ limit is defined as $c\rar \infty$ with $\Delta,q^{(3)}$ fixed, or alternatively  
\bea 
b\rar 0 
\scs 
\al = -bj\scs j\ {\rm fixed}\ .
\label{baj}
\eea
This is sometimes called the light asymptotic limit. 
In the large $c$ limit, the $W_3$ algebra reduces to $s\ell_3$, and the vector $j$ is the spin of an $s\ell_3$ representation. This spin is related to the limits $\Delta=\underset{b\rar 0}{\lim} \Delta_{-bj}$ and $q^{(3)} = \underset{b\rar 0}{\lim}q^{(3)}_{-bj}$ by 
\bea
\Delta =  r+s \scs q^{(3)} = r-s \ ,
\label{dq}
\eea
where $r$ and $s$ are the components of the spin $j$, see eq. (\ref{rjsj}).
Let us define the large $c$ four-point conformal blocks,
\bea
{\cal F}_{g,g'|j_s}(j_i|z_i) = \underset{b\rar 0}{\lim}\ {\cal G}_{g,g'|-bj_s}(c|-bj_i|z_i)\ .
\eea
In analogy with the case of Virasoro conformal blocks, the large $c$ conformal blocks of the $W_3$ algebra can be computed as special cases of $s\ell_3$-invariant functions. We claim that an isospin variable $Z=(w,x,y)$ of an $s\ell_3$-invariant function must then be of the type $Z=\vec{z}$ where we define
\bea
\vec{z} = (\tfrac12 z^2 , z ,z )\ .
\label{zz}
\eea
This relation between the isospin $Z$ and the worldsheet position $z$ comes from the following identities, which hold for any function ${\cal E}(Z)$:
\bea
\left. D_{(j,Z)}(h^1+h^2){\cal E}(Z) \right|_{Z=\vec{z}} &=& -D_{(\Delta,z)}(L_0){\cal E}(\vec{z}) \ ,
\\
\left. D_{(j,Z)}(e^1+e^2){\cal E}(Z) \right|_{Z=\vec{z}} &=& -\tfrac12 D_{(\Delta,z)}(L_{1}){\cal E}(\vec{z}) \ ,
\\
\left. D_{(j,Z)}(f^1+f^2){\cal E}(Z) \right|_{Z=\vec{z}} &=& D_{(\Delta,z)}(L_{-1}){\cal E}(\vec{z}) \ ,
\eea
where the $s\ell_2$ differential operators $D_{(\Delta,z)}(t^a)$ were defined in eq. (\ref{ddd}), and the $s\ell_3$ operators $D_{(j,Z)}(t^a)$ in eq. (\ref{dh})-(\ref{df}). These identities show that
the principally embedded $s\ell_2$ subalgebra of $s\ell_3$ with generators $(h^1+h^2,e^1+e^2,f^1+f^2)$ can be identified with the $s\ell_2$ algebra of global conformal transformations. (See also \cite{bw91}.)
Therefore, a large $c$ four-point block is a special case of a four-point $s\ell_3$-invariant function (\ref{egg}),
\bea
{\cal F}_{g,g'|j_s}(j_i|z_i) = {\cal N}\int_C dZ_s\ {\cal E}_g(j_1,j_2,j_s|\vec{z}_1,\vec{z}_2,Z_s) {\cal E}_{g'}(j_s^{*\om},j_3,j_4|Z_s,\vec{z}_3,\vec{z}_4)\ ,
\label{fgg}
\eea
where the maximal Weyl reflection $j\rar j^*$ and the Dynkin diagram automorphism $j\rar j^\om$ were defined in eqs. (\ref{jjs}) and (\ref{jje}), and the three-point invariant ${\cal E}_g$ was given in eq. (\ref{egc}).
The normalization factor ${\cal N}$ and the integration domain $C$ for $Z_s\in \C^3$ are determined by the condition (\ref{fzo}). Other choices of integration domains in eq. (\ref{fgg}) would lead to linear combinations of six conformal blocks whose spins $j_s$ are related by Weyl transformations (\ref{rsr}). 
Notice that the condition (\ref{fzo}) of analyticity and normalization of conformal blocks also constrains the multiplicity functions $g,g'$. Applying that condition to a three-point invariant function ${\cal E}_g(j_1,j_2,j_s|\vec{z}_1,\vec{z}_2,Z_s)$ leads to the conditions
\bea
g(1)=2^{\frac13(r+2s-2r_1-s_1-2r_2-s_2)} \scs g(\theta)\ {\rm is\ analytic\ near}\ \theta =1\ .
\eea

We will now focus on a large $c$ four-point block ${\cal F}_{j_s}(j_i|z_i)$ such that the fields with numbers $2,3$ are semi-degenerate of the first kind, so that the multiplicity functions $g,g'$ disappear and the components $r_2,r_3$ of the spins $j_2,j_3$ vanish. 
The assumptions $Z_1=\vec{z}_1$ and $Z_2=\vec{z}_2$ lead to simplifications in the combinations $\rho_{12}$ (\ref{rho}) and $\chi_{123}$ (\ref{chi}), 
\bea
\rho_{12}(\vec{z}_1,\vec{z}_2) &=& \tfrac12 z_{12}^2 \ ,
\\
\chi_{123}(\vec{z}_1,\vec{z}_2,Z_3) &=& \tfrac12 z_{12}\left(z_1z_2-y_3z_1-y_3z_2 +2x_3y_3-2w_3\right)\ ,
\eea
so that the relevant three-point invariants eq. (\ref{frz}) become
\begin{multline}
{\cal E}(j_1,j_2,j_3|\vec{z}_1,\vec{z}_2,Z_3) = 2^{-J-r_3+s_1+s_2} 
z_{12}^{\Delta_3-\Delta_1-\Delta_2} 
\\ \times 
\left(z_1z_2-y_3z_1-y_3z_2 +2x_3y_3-2w_3\right)^{-J} \left(y_3x_3-y_3z_1-w_3+\tfrac12 z_1^2\right)^{J-s_3}
\\ \times
\left(w_3-x_3z_2+\tfrac12 z_2^2\right)^{-J-r_3+s_1} \left(w_3-x_3z_1+\tfrac12 z_1^2\right)^{J-s_1}  \ , \qquad (r_2=0) \ .
\label{ez}
\end{multline}
where 
the conformal dimensions $\Delta_i$ are associated to the spins $j_i$ as in eq. (\ref{dq}).
Then the formula (\ref{fgg}) implies that $z^{\Delta_1+\Delta_2} {\cal F}_{j_s}(j_i|0,z,1,\infty)$ depends on only four combinations of the six nonvanishing components of the spins $j_1,j_2,j_3,j_4$, namely
\bea
r_1-s_2 \scs s_1 \scs r_4-s_3 \scs s_4\ .
\label{fs}
\eea
It is actually convenient to use the following four combinations, where we call $(r,s)$ the components of $j_s$,
\bea
\begin{array}{lcl}
\al = \tfrac13(s_3+s_4+s-r_4-r)\ ,  & \quad  & \beta = \tfrac13(s_1+s_2-s-r_1+r) \ , 
\\
\g = \al-s_4+r \ , & & \delta = \beta-s_1+s\ .
\end{array} 
\label{abcd}
\eea 
The integral in eq. (\ref{fgg}) can be expanded near $z_1=z_2$, see Appendix \ref{appder}. This leads to the expansion of the large $c$ four-point block ${\cal F}_{j_s}(z)={\cal F}_{j_s}(j_i|0,z,1,\infty)$ near $z=0$,
\bea
{\cal F}_{j_s}(z) &=& z^{-\Delta_{12}^s} \sum_{n,i,j=0}^\infty \frac{z^{2n+i+j}}{n!i!j!(r+s-1)_n} \frac{(\beta)_i(-\beta+r)_n(\g)_{n+i}}{(r)_{n+i}} \frac{(\al)_j(-\al+s)_n(\delta)_{n+j}}{(s)_{n+j}} \ ,
\label{fsss}
\\
&=& z^{-\Delta_{12}^s}\sum_{n=0}^\infty \frac{z^{2n}}{n!}\frac{(r-\beta)_n(\g)_n (s-\al)_n(\delta)_n}{(r)_n(s)_n(r+s-1)_n} 
\nn
\\ && \hspace{3cm} \times
F(\beta,\g+n,r+n,z) F(\al,\delta+n,s+n,z)\ ,
\eea
where the notation $(t)_n$ was defined in eq. (\ref{tn}). The second form of this expression is obtained by performing the sums over $i$ and $j$, and can be helpful in numerical computations.
A similar expression can be obtained if the fields with numbers $2,3$ are semi-degenerate of the second kind (instead of the first kind), by exchanging the two components $r$ and $s$ of each spin. 

There are five special cases where the block ${\cal F}_{j_s}(z)$ (\ref{fsss}) reduces to a hypergeometric function:
\begin{enumerate}
\item \underline{Case $\al=\beta=0$:} In this case $(\beta)_i=\delta_{i0}$ and $(\al)_j=\delta_{j0}$, so that
\bea
{\cal F}_{j_s}(z) = z^{-\Delta_{12}^s} F(\g,\delta,r+s-1,z^2) \ .
\label{fabz}
\eea
This shows that blocks can have a singularity at $z=-1$, in addition to the physical singularities at $z=0,1,\infty$ which appear when two of the fields at $z_1,z_2,z_3,z_4$ come together.
\item \underline{Case $\g=0$:} In this case $(\g)_{n+i} = \delta_{n+i,0}$ and we have
\bea
{\cal F}_{j_s}(z) = z^{-\Delta_{12}^s}F(\al,\delta,s,z) \ .
\label{fgz}
\eea
\item \underline{Case $\delta=0$:} In this case $(\delta)_{n+j} = \delta_{n+j,0}$ and we have
\bea
{\cal F}_{j_s}(z) = z^{-\Delta_{12}^s}F(\beta,\gamma,r,z)\ .
\label{fdz}
\eea
\item \underline{Case $\al=s$:} In this case $(s-\al)_n=\delta_{n,0}$ and we have 
\bea
{\cal F}_{j_s}(z) = z^{-\Delta_{12}^s}(1-z)^{-\delta}F(\beta,\gamma,r,z)\ .
\label{fas}
\eea
\item \underline{Case $\beta=r$:} In this case $(r-\beta)_n=\delta_{n,0}$ and we have
\bea
{\cal F}_{j_s}(z) = z^{-\Delta_{12}^s}(1-z)^{-\g}F(\al,\delta,s,z)\ .
\label{fbr}
\eea
\end{enumerate}

\zeq\section{Differential equation and critical exponents for large $c$ conformal blocks \label{appdc}}

\subsection{Case of Virasoro conformal blocks}

The large $c$ limit ${\cal F}_{\Delta_s}(\Delta_i|z_i)$ of a Virasoro four-point block obeys a second-order hypergeometric differential equation, and we now explain how to deduce this equation from the integral expression (\ref{fds}) for ${\cal F}_{\Delta_s}(\Delta_i|z_i)$. The large $c$ block is indeed a function of four variables $z_i$, which obeys the three equations (\ref{tlll}) of global conformal invariance. In addition, the three-point invariant ${\cal E}(\Delta_1,\Delta_2,\Delta_s|z_1,z_2,z_s)$ which appears in the integral expression (\ref{fds}) also obeys these equations, and together with the relation (\ref{cdd}) for the quadratic Casimir $C_2(\Delta)$ this implies 
\bea
g_{ab}(D_{(\Delta_1,z_1)}+D_{(\Delta_2,z_2)})(t^a)(D_{(\Delta_1,z_1)}+D_{(\Delta_2,z_2)})(t^b) {\cal F}_{\Delta_s}(\Delta_i|z_i) = C_2(\Delta_s) {\cal F}_{\Delta_s}(\Delta_i|z_i) \ , 
\eea
where the differential operators $D_{(\Delta,z)}(t^a)$ are defined in eq. (\ref{ddd}). Thus ${\cal F}_{\Delta_s}(\Delta_i|z_i)$ obeys four differential equations, and ${\cal F}_{\Delta_s}(\Delta_i|0,z,1,\infty)$ obeys one differential equation, which turns out to be the hypergeometric equation, whose solution (subject to the condition (\ref{fzo})) we wrote in eq. (\ref{zf}).

The critical exponents of the hypergeometric equation are known, and we deduce the critical exponents $\lambda_i$ of $z^{\Delta_1+\Delta_2}{\cal F}_{\Delta_s}(\Delta_i|0,z,1,\infty)$ at the three singularities $z=0,1,\infty$:
\bea
\begin{array}{|c||c|c|c|}
\hline
exponent & 0 & 1 & \infty
\\
\hline
\hline
\lambda_1 & \Delta_s & 0 & \Delta_1-\Delta_2
\\
\hline
\lambda_2 & \Delta_s^* & \Delta_1+\Delta_4-\Delta_2-\Delta_3 & \Delta_4-\Delta_3
\\
\hline
\end{array}
\label{let}
\eea
The number $\lambda_2^{(0)} = \Delta_s^* = 1-\Delta_s$ is not really a critical exponent of the block itself, rather it corresponds to another block with another $s$-channel dimension $\Delta_s^*$. Notice that the exponents at $1$ and $\infty$ are $\Delta_s$-independent; conformal blocks can be expected to behave so simply at these singularities only in the large $c$ limit. The transformations of the hypergeometric functions can be used to rewrite ${\cal F}_{\Delta_s}(\Delta_i|0,z,1,\infty)$ as a combination of two functions with simple monodromy at say $z=1$, but these two functions are not themselves conformal blocks in another channel.

\subsection{Differential equation for large $c$ $W_3$ conformal blocks}

That our large $c$ four-point conformal block ${\cal F}_{j_s}(j_i|z_i)$ (\ref{fsss}) with two semi-degenerate fields obeys a differential equation follows from a simple counting of variables and equations. We consider first the corresponding $s\ell_3$ four-point invariant function ${\cal E}_{j_s}(j_i|Z_i)$, which depends on $12$ isospin variables, namely the components of $Z_1,Z_2,Z_3,Z_4$, and is such that ${\cal F}_{j_s}(j_i|z_i)={\cal E}_{j_s}(j_i|\vec{z}_i)$ where $\vec{z}$ is defined in eq. (\ref{zz}). There are three types of equations for ${\cal E}_{j_s}(j_i|Z_i)$:
\begin{enumerate}
\item The $s\ell_3$ symmetry condition (\ref{sde}) yields $8$ equations.
\item As two fields are semi-degenerate, we have two equations $d^{(1)}_{Z_2} {\cal E}_{j_s}(j_i|Z_i) = d^{(1)}_{Z_3} {\cal E}_{j_s}(j_i|Z_i)= 0 $, where $d^{(1)}_Z$ was defined in eq. (\ref{dzdz}).
\item The $s\ell_3$ symmetry condition for the three-point invariant ${\cal E}_g(j_1,j_2,j_s|Z_1,Z_2,Z_s)$ which appears in the integral formula (\ref{egg}) will yield two more equations. The $s\ell_3$ symmetry condition (\ref{sde}) applied to ${\cal E}_g(j_1,j_2,j_s|Z_1,Z_2,Z_s)$ indeed implies
\bea
g_{ab}(D_{(j_1,Z_1)}+D_{(j_2,Z_2)})(t^a)(D_{(j_1,Z_1)}+D_{(j_2,Z_2)})(t^b) {\cal E}_{j_s}(j_i|Z_i) = C_2(j_s) {\cal E}_{j_s}(j_i|Z_i) \ ,
\label{gfcf}
\eea
\begin{multline}
d_{abc}(D_{(j_1,Z_1)}+D_{(j_2,Z_2)})(t^a)(D_{(j_1,Z_1)}+D_{(j_2,Z_2)})(t^b)(D_{(j_1,Z_1)}+D_{(j_2,Z_2)})(t^c) {\cal E}_{j_s}(j_i|Z_i)
\\
= -C_3(j_s) {\cal E}_{j_s}(j_i|Z_i) \ ,
\label{dfcf}
\end{multline}
using the equations (\ref{gi}) and (\ref{di}) which involve the 
the Casimir numbers $C_2(j_s)$ and $C_3(j_s)$.
\end{enumerate}
The function ${\cal E}_{j_s}(j_i|Z_i)$ of $12$ variables therefore obeys $12$ partial differential equations. Now the function ${\cal F}_{j_s}(j_i|z_i)$ can be written in terms of a function of just one variable (the cross-ratio of $z_1,z_2,z_3,z_4$), and this function obeys one differential equation. The order of the differential equation can be guessed to be six, the order of the Weyl group of $s\ell_3$. This is because our differential equations (\ref{gfcf}) and (\ref{dfcf}) depend on the spin $j_s$ through the Weyl invariants $C_2(j_s)$ and $C_3(j_s)$. Given a solution, Weyl reflections of $j_s$ therefore provide five other solutions. 

Let us explain how the differential equation for ${\cal F}_{j_s}(j_i|z_i) $ can be derived in principle. We will not perform the derivation to the end, as the resulting equation would be too complicated to be useful. We will stop at the partial differential equations for $ {\cal E}_{j_s}(j_i|Z_i)$, which imply the equation for ${\cal F}_{j_s}(j_i|z_i) $ and are much simpler. It is from these partial differential equations that we will derive interesting information like the critical exponents of $ {\cal F}_{j_s}(j_i|z_i)$. To begin with, the first ten equations allow us to rewrite $ {\cal E}_{j_s}(j_i|Z_i)$ in terms of a function of two variables $\hat{\cal E}_{j_s}(j_i|U,V)$,
\bea
{\cal E}_{j_s}(j_i|Z_i) = Q(j_i|Z_i) \hat{\cal E}_{j_s}(j_i|U,V)  \ ,
\label{ftf}
\eea
where the cross-ratios $U,V$ are solutions of our ten equations when all spins are taken to zero,
\bea
U  = \frac{\rho_{34}}{\rho_{24}}\frac{\chi_{412}}{\chi_{431}} \scs V =\frac{\rho_{21}}{\rho_{31}}\frac{\chi_{341}}{\chi_{421}} \ ,
\label{uv}
\eea
and the prefactor $Q(j_i|Z_i)$ is the product of two three-point invariants of the type (\ref{frz}) with one spin set to zero in each invariant (thereby imitating the prefactor $P(\Delta_i|z_i)$ (\ref{oz}) of $s\ell_2$-invariant functions), 
\bea
Q(j_i|Z_i) &=&  {\cal E}(j_1,j_2,0|Z_1,Z_2,Z_4) {\cal E}(0,j_3,j_4|Z_1,Z_3,Z_4) \ ,
\label{tff}
\\
&=&
\chi_{124}^{-j_{12}}\chi_{134}^{-j_{34}} \rho_{41}^{j_{12}+j_{34}-s_4} \rho_{14}^{j_{12}+j_{34}-s_1} \rho_{21}^{-j_{12}-r_1}\rho_{31}^{-j_{34}+s_4} \rho_{34}^{-j_{34}-r_4} \rho_{24}^{-j_{12}+s_1} \ ,
\eea
where we defined
\bea
j_{12} = (h_2,j_1+j_2) = \tfrac13(s_1+s_2-r_1) \ , \ j_{34} = (h_2,j_3+j_4) = \tfrac13(s_3+s_4-r_4) \ .
\label{jj}
\eea
We have $Q(j_i|\vec{z}_i)=P(\Delta_i|z_i)$ where $P(\Delta_i|z_i)$ was defined in eq. (\ref{oz}) and $\Delta_i$ in eq. (\ref{dq}), and together with the expression (\ref{fog}) for ${\cal F}_{j_s}(j_i|z_i)$ in terms of  ${\cal F}_{j_s}(z)={\cal F}_{j_s}(j_i|0,z,1,\infty)$  this implies
\bea
z^{\Delta_1+\Delta_2}{\cal F}_{j_s}(z) = \hat{\cal E}_{j_s}(j_i|z,z)\ .
\label{fe}
\eea
The two equations (\ref{gfcf}) and (\ref{dfcf}) amount to two partial differential equations for $\hat{\cal E}_{j_s}(j_i|U,V)$, which we computed with the help of the free mathematical software Sage. The equations are of the type $E_2 \hat{\cal E}_{j_s}(j_i|U,V) = E_3 \hat{\cal E}_{j_s}(j_i|U,V) = 0$, where the differential operators $E_2$ and $E_3$ are 
\begin{multline}
 E_2 = D_U^2+D_V^2-D_UD_V-D_U-D_V-\tfrac12 C_2(j_s)
\\ 
+U(D_U+j_{12}-s_1)(D_V-D_U-j_{34}) 
+V(D_V+j_{34}-s_4)(D_U-D_V-j_{12}) 
\\
-UV(D_U+j_{12}-s_1)(D_V+j_{34}-s_4)
\ ,
\label{ee}
\end{multline}
\begin{multline}
E_3 = (D_V-D_U)(D_V-1)(D_U-1) -\tfrac16 C_3(j_s)
\\
 - U(D_U+j_{12}-s_1)(D_V-D_U-j_{34})(D_V-1) 
+V(D_V+j_{34}-s_4)(D_U-D_V-j_{12})(D_U-1) 
\\
+UV(D_U+j_{12}-s_1)(D_V+j_{34}-s_4)(D_V-D_U+j_{12}-j_{34})
 \ ,
\label{eee}
\end{multline}
where we defined $D_U = U\pp{U}$ and $D_V = V\pp{V}$. (The combinations $j_{12},j_{34}$ of the components $r_i,s_i$ of the spins $j_i$ were defined in eq. (\ref{jj}).) The differential operators $E_2$ and $E_3$ commute, as guaranteed by their origin in the $s\ell_3$-invariant differential operators which appear in eqs. (\ref{gfcf}) and (\ref{dfcf}).

Let us sketch how a sixth-order differential equation for $\hat{\cal E}_{j_s}(j_i|z,z)$ is obtained from the two partial differential equations $E_2$ and $E_3$ for $\hat{\cal E}_{j_s}(j_i|U,V)$. We cannot directly set $U=V$ in $E_2$ and $E_3$, because these differential operators do not keep the line $\{U=V\}$ invariant. To cure this problem,
we take linear combinations of $E_2$ and $E_3$ with differential operators as coefficients, so as to eliminate $D_U-D_V$ while keeping the derivative $D_U+D_V$ along $\{U=V\}$. This yields a differential operator of the type $E_6 = \sum_{i=0}^6 c_i(U,V) (D_U+D_V)^i$ such that $E_6 \hat{\cal E}_{j_s}(j_i|U,V)=0$, and we thus have $\left[\sum_{i=0}^6 c_i(z,z) (2z\pp{z})^i \right] z^{\Delta_1+\Delta_2}{\cal F}_{j_s}(z) =0$. 

The resulting differential equation is however too complicated to be useful. We were able to compute it explicitly (with the help of a computer) only in special cases when some parameters $j_i,j_s$ vanish. Even so, the equation is rather complicated, and we do not display it. We will however discuss its singularities and the corresponding critical exponents. Knowing the critical exponents at a given singularity is equivalent to knowing the leading term of the differential equation $E_6$  near that singularity. To derive this, 
the algorithm for obtaining the differential equation $E_6$ from $E_2$ and $E_3$ can be applied to the first few leading terms of $E_2,E_3$ near the singularity. (Keeping one term of each equation is in general not enough, except at the singularity $z=0$ as we shall see.)

\subsection{Singularities and critical exponents of large $c$ $W_3$ conformal blocks}

A four-point correlation function $\la \prod_{i=1}^4 V_{\al_i}(z_i)\ra $ in $s\ell_3$ conformal Toda theory (or actually in any conformal field theory) is expected to have singularities at $z_i=z_j$, which in terms of the cross-ratio amounts to $z=0,1,\infty$. A conformal block like ${\cal G}_{g,g'|\al_s}(c|\al_i|z_i)$, and its large $c$ limit ${\cal F}_{g,g'|j_s}(j_i|z_i)$, is therefore also expected to be singular at these points. However, nothing in principle excludes the existence of extra singularities in conformal blocks, and in the case when two fields are semi-degenerate, we will indeed find that ${\cal F}_{j_s}(j_i|z_i)$ has an unexpected singularity at $z=-1$, as we already noticed in a special case (\ref{fabz}). 

While it is not clear to us why this singularity appears, we can at least explain why the point $z=-1$ is special. The set $\{0,1,\infty\}$ of the physical singularities is invariant under a set of six $PSL_2(\Z)$ transformations $z\rar (z,1-z,\frac{1}{z},\frac{1}{1-z},\frac{z}{z-1},1-\frac{1}{z})$. But in our correlation functions the fields $2$ and $3$ are semi-degenerate, and the only nontrivial transformation which does not mix them with the other fields is $z\rar \frac{1}{z}$. The point $z=-1$ is characterized as the nontrivial fixed point of that transformation. Notice that this argument is specific neither to $s$-channel conformal blocks (as opposed to blocks in other channels or to correlation functions), nor to the large $c$ limit (as opposed to generic values of $c$). 

But let us first comment on the singularity at $z=0$. In the limit $U,V\rar 0$, the partial differential equations $E_2$ (\ref{ee}) and $E_3$ (\ref{eee}) for $\hat{\cal E}_{j_s}(j_i|U,V)$ are reduced to their respective first lines. 
Let look for solutions of the type
\bea
\hat{\cal E}_{j_s}(j_i|U,V)= U^\mu V^\nu \sum_{m,n=0}^\infty c_{m,n} U^n V^m\ , \qquad ({\rm assuming\ } c_{0,0}=1)\ ,
\label{euv}
\eea
for some exponents $(\mu,\nu)$. Using the expressions (\ref{kk}) and (\ref{kkk}) for $C_2(j_s)$ and $C_3(j_s)$ respectively, we find six solutions which correspond to the pairs $(\mu,\nu)$ 
such that $j_s=-\nu e_1-\mu e_2$ up to Weyl reflections. We adopt the solution $(\mu,\nu) = (\frac{2s+r}{3},\frac{2r+s}{3})$ where $(r,s)$ are the components of $j_s$. 
The equation $E_2 \hat{\cal E}_{j_s}(j_i|U,V) = 0$ leads to a recursion relation for the coefficients $c_{m,n}$,
\begin{multline}
(m^2+n^2-mn-m-n+sn+rm) c_{m,n} +(m-1+\g)(n-m+1-\beta)c_{m-1,n}
\\
 + (n-1+\delta)(m-n+1-\al)c_{m,n-1} -(n-1+\delta)(m-1+\g)c_{m-1,n-1} =0\ ,
\end{multline}
where the combinations $\al,\beta,\g,\delta$ of spin components were defined in eq. (\ref{abcd}).
This relation has a unique solution such that $c_{0,0}=1$ (assuming $c_{m,n}=0$ unless $m,n\geq 0$), which is
\bea
c_{m,n}= \frac{(\g)_m(\delta)_n}{(r)_m(s)_n} \sum_{h=0}^{\min(m,n)} \frac{(\beta)_{m-h}(r-\beta)_h(\al)_{n-h}(s-\al)_h}{h!(m-h)!(n-h)!(r+s-1)_h}\ .
\eea
So, once the critical exponents $(\mu,\nu)$ are deduced from $E_2$ and $E_3$, the equation $E_2$ is enough to determine the solution uniquely. 
The equation $E_3\hat{\cal E}_{j_s}(j_i|U,V) = 0$ leads to another recursion relation, which however has the same solution. Setting $U=V=z$ in $\hat{\cal E}_{j_s}(j_i|U,V)$ as in eq. (\ref{fe}), we recover the expression (\ref{fsss}) for ${\cal F}_{j_s}(z)$.

It is less straightforward to compute the critical exponents of $z^{\Delta_1+\Delta_2}{\cal F}_{j_s}(z)$ at $z=1,\infty,-1$ than at $z=0$, and we present only the results. The function $z^{\Delta_1+\Delta_2}{\cal F}_{j_s}(z)$ obeys a differential equation of order six, and therefore has six exponents at each singularity, which we number arbitrarily: 
\footnote{The exponents which we write are valid when the values of the spins are generic. In the special cases when two of the exponents at a given singularity coincide, complications can occur, including the appearance of logarithmic terms in the expansion of ${\cal F}_{j_s}(z)$.}
\bea
\begin{array}{|c||c|c|c|c|}
\hline
exponent & 0 & \infty & 1 & \ \ -1\ \  
\\
\hline
\hline
\lambda_1 & r+s & r+s-\g-\delta-1 & r+s-\g-\delta-1 & \lambda
\\
\hline
\lambda_2 & 1+r & r+s-\g-\delta & r+s-\g-\delta & 0 
\\
\hline
\lambda_3 & 1+s & r+s-\al-2\g & 0 & 1 
\\
\hline
\lambda_4 & 3-r & r+s-\delta-2\beta & r+s-\al-\beta-\g-\delta & 2 
\\
\hline
\lambda_5 & 3-s & r+\al-\beta-\delta & s-\al-\delta & 3
\\
\hline
\lambda_6 & 4-r-s & s+\beta-\al-\g & r-\beta-\g & 4 
\\
\hline
\end{array}
\label{tet}
\eea
where the nontrivial critical exponent at $z=-1$ is
\bea
\lambda = s_1+s_4-1 = r+s+\al+\beta -\g-\delta-1 \ .
\label{lrs}
\eea
These critical exponents of $z^{\Delta_1+\Delta_2}{\cal F}_{j_s}(z)$ can be compared with those of conformal blocks of the Virasoro algebra (\ref{let}). Notice that the critical exponents at a regular point are $(0,1,2,3,4,5)$, so the point $-1$ is almost regular in that only one exponent is not an integer. The sum of all exponents (with a minus sign for the exponents at $\infty$) is $\sum_{i=1}^6 (\lambda_i^{(0)} -\lambda_i^{(\infty)} + \lambda_i^{(1)} + \lambda_i^{(-1)}) = 21$. 

In the five special cases at the end of Subsection \ref{sscbl}, the blocks reduce to hypergeometric functions, and at each singularity we recover two exponents out of six. 
Most of the critical exponents (\ref{tet}) predicted by the differential equation can thus be confirmed in these special cases.

Finally, the existence of the surprising singularity at $z=-1$, and the value of the critical exponent $\lambda$, can be confirmed using the integral representation (\ref{fgi}) of the conformal block. After some manipulations which are rather straightforward, we indeed find that the block has the asymptotic behaviour (if $\lambda<0$)
\bea
{\cal F}_{j_s}(z) \underset{z\rar -1}{\sim} \frac{\G(r)\G(s)\G(r+s-1) \G(-\lambda)}{\G(\g)\G(\delta)\G(s-\al)\G(r-\beta)} 2^{\lambda} (z+1)^{\lambda}\ ,
\eea
As expected, the coefficient of $(z+1)^\lambda$ vanishes in the four special cases where the singularity at $z=-1$ disappears, see eq. (\ref{fgz})-(\ref{fbr}).

\section{Conclusion \label{sconc}}

\subsection{Comparison with the combinatorial expansion}

A combinatorial expansion is proposed in \cite{fl11} for four-point conformal blocks of the $W_3$ algebras, such that two fields are semi-degenerate and therefore no infinite fusion multiplicities are present. If the fields with numbers $2,3$ have momenta along the weight $h_1$, that is $\bla \al_2 = r_2 h_1 \\ \al_3 = r_3 h_1 \ela$, then the four-point block ${\cal G}_{\al_s}(c|\al_i|0,z,1,\infty)$ reads
\bea
{\cal G}_{\al_s}(c|\al_i|0,z,1,\infty) &=&
(1-z)^{r_3(\frac{1}{3} r_2-q)} z^{-\Delta_{12}^s} \sum_{\vec{\lambda}} z^{|\vec{\lambda}|} \frac{F_{\vec{0},\vec{\lambda}}(\al_4^\om,\al_s,r_3) F_{\vec{\lambda},\vec{0}}(\al_s,\al_1,r_2)}{F_{\vec{\lambda},\vec{\lambda}}(\al_s,\al_s,0)}\ .
\label{guw}
\eea
where the function $F_{\vec{\lambda},\vec{\lambda}'}(\al,\al',r)$ is defined as
\bea
F_{\vec{\lambda},\vec{\lambda}'}(\al,\al',r)&=& \prod_{i,j=1}^3 \left(\prod_{s\in \lambda_i'}\left[ (h_j,\al-Q)-(h_i,\al'-Q) -\tfrac{1}{3}r - bl_{\lambda_j}(s)+b^{-1}(a_{\lambda'_i}(s)+1)\right] \right.
\nn
\\
&& \times \left. \prod_{s\in \lambda_j}\left[ (h_j,\al-Q)-(h_i,\al'-Q) -\tfrac{1}{3}r +b(l_{\lambda_i'}(s)+1)-b^{-1}a_{\lambda_j}(s)\right]\right)\ .
\eea
Besides the "$W_3$ notations" $\Delta,\al,Q,q,b$ introduced in Subsection \ref{sscbl} and the definition (\ref{jje}) of the Dynkin diagram automorphism $j\rar j^\om$, these formulas use notations for Young diagrams which we now review. (See \cite{aflt10} for more details.) The sum in eq. (\ref{guw}) is over triples $\vec{\lambda} = (\lambda_1,\lambda_2,\lambda_3)$ of Young diagrams. Each diagram is a collection of boxes, and each box $s$ has an arm length $a_\lambda(s)$ and leg length $l_\lambda(s)$ relative to a diagram $\lambda$, which are positive if $s\in \lambda$. The triple $\vec{0}=(0,0,0)$ is the set of three empty diagrams.

We have checked that the large $c$ limit (\ref{baj}) of the block (\ref{guw}) agrees with the independently derived prediction eq. (\ref{fsss}) up to the order $z^5$, modulo the exchange of the components $r$ and $s$ of the spins due to the use of different kinds of semi-degenerate fields. The agreement is rather non-trivial, because individual terms of the sum over Young diagrams can have spurious poles (as functions of the components of $\al_s$), and these spurious poles cancel when the sum is performed.

\subsection{Concluding remarks}

In this article, we have given an integral formula (\ref{fgg}) for the large $c$ limits of arbitrary $W_3$ conformal blocks on the sphere. This formula is a special case of an $s\ell_3$-invariant function, where the isospin variables take special values determined by the positions of the fields. This result implies that the large $c$ conformal block depends nontrivially on only eight combinations of the ten components of the spins $(j_s,j_1,j_2,j_3,j_4)$, in the same way as the large $c$ Virasoro conformal block (\ref{zf}) depends nontrivially on only three combinations of the five conformal dimensions.

We have thus shown how to take infinite fusion multiplicities into account in this limit. This might be helpful for solving the problem of the infinite fusion multiplicities in general, and allow us to
deal with arbitary conformal blocks in conformal field theories with $W_N$ symmetries. So far, we know combinatorial expansions only for blocks with no fusion multiplicities. For fusion multiplicities to be absent, it is necessary to restrict the momenta of the fields, such that they are all almost fully degenerate except two of them. (See also \cite{fl11}.) While we can deal with large $c$ blocks in general, imposing such restrictions brings important simplifications. 
Thus we studied the detailed properties a certain class
of $W_3$ large $c$ four-point conformal blocks with two semi-degenerate fields, and in particular we derived their series expansion (\ref{fsss}).
We found that, for generic values of the parameters, such blocks have a singularity at $z=-1$, in addition to the expected singularities at $z=0,1,\infty$. We believe that the singularity at $z=-1$ is absent for non-infinite values of the central charge $c$. It would be interesting to confirm this expectation, and to understand how the singularity disappears for finite values of $c$.

We expect that our results can be generalized to $s\ell_N$-invariant functions and $W_N$ conformal blocks in the large $c$ limit. In particular, we expect the large $c$ limit of a $W_N$ four-point conformal block with two almost fully degenerate fields to obey a differential equation of order $N!$, as we observed in the cases $N=2$ and $N=3$ in Section \ref{appdc}.


\appendix

\zeq\section{Quantum mechanics of a point particle on $SL_N(\C)$ \label{appqm}}

For any integer $N\geq 2$, the algebra $W_N$ is the symmetry algebra of a CFT called $s\ell_N$ conformal Toda theory. In the case $N=2$ for example, the Virasoro ($W_2$) algebra is the symmetry algebra of Liouville ($s\ell_2$ conformal Toda) theory. The functional integral representation of the correlation functions of $s\ell_N$ conformal Toda theory can be used for studying their large $c$ limits, which turn out to be correlation functions of the quantum mechanics of a point particle on $SL_N(\C)$. (See for instance \cite{fl07c}.) This will provide some justification for our identification of large $c$ conformal blocks with $s\ell_N$-invariant functions. We will start with 
the case of the point particle on $SL_2(\C)$, before dealing with the technically more complicated case of $SL_3(\C)$.

\subsection{Point particle on $SL_2(\C)$ \label{saqm}}

Let us call $V_\Delta(z,\bz)$ a primary vertex operator of Liouville theory. This depends on the complex coordinates $(z,\bz)$ of a point on the complex plane, and on the conformal dimension $\Delta$ of the corresponding Virasoro representation. The functional integral representation of a correlation function $\la \prod_{i=1}^n V_{\Delta_i}(z_i,\bz_i) \ra$ of $n$ such vertex operators leads to the large $c$ limit
\bea
\underset{c\rar \infty}{\lim} 
\la \prod_{i=1}^n V_{\Delta_i}(z_i,\bz_i) \ra = \int_{SL_2(\C)}dg\ \prod_{i=1}^n \Phi^{\Delta_i}_{z_i,\bz_i}(g) \ ,
\label{oip}
\eea
where $\Phi^{\Delta}_{z,\bz}(g)$ is the classical limit of the field $V_\Delta(z,\bz)$, evaluated on a solution of the Liouville equation labelled by $g\in SL_2(\C)$,
\bea
\Phi^\Delta_{z,\bz}(g) = \left[ v_z g v_{\bz}^T \right]^{-2\Delta} \scs v_z = (1,z)\ .
\label{pv}
\eea
(Real solutions of the Liouville equation actually correspond to Hermitian matrices $g$. We will neglect this subtlety.)
The functions $\Phi^\Delta_{z,\bz}(g)$ and $v_z$ have a simple behaviour under $s\ell_2$ tranformations of the matrix $g$:
\bea
v_zt^a = D_{(-\frac12,z)}(t^a)v_z \scs 
\Phi^\Delta_{z,\bz}((1+\e t^a)g) = (1+\e D_{(\Delta,z)}(t^a)) \Phi^\Delta_{z,\bz}(g) + O(\e^2) \ ,
\label{pded}
\eea
where the differential operators $D_{(\Delta,z)}(t^a)$ were introduced in eq. (\ref{ddd}), and 
the generators $(t^a)=(L_1,L_0,L_{-1})$ of $s\ell_2$ are realized as the matrices 
\bea
L_{-1} = \bsm 0 & -1 \\ 0 & 0 \esm \scs L_0 = \tfrac12\bsm 1 & 0 \\ 0 & -1 \esm \scs L_{1} = \bsm 0 & 0 \\ 1 & 0 \esm\ .
\eea
Now the differential operators $D_{(\Delta,z)}(t^a)$ obey the relation
\bea
g_{ab} D_{(\Delta,z)}(t^a)D_{(\Delta,z)}(t^b) = C_2(\Delta)\ ,
\label{cdd}
\eea
where we defined
\bea
g^{ab} = 2\Tr t^a t^b \scs C_2(\Delta)=\Delta(\Delta-1)\ .
\label{gcdd}
\eea
Together with eq. (\ref{pded}), this implies that $\Phi^\Delta_{z,\bz}(g)$ is an eigenvector of the $s\ell_2$-invariant Laplacian on $SL_2(\C)$. Actually,
the functions $\Phi^\Delta_{z,\bz}(g)$ provide a basis of functions of $SL_2(\C)$, whose completeness can be expressed as the decomposition of the Dirac delta function $\delta(g,g')$,
\bea
\delta(g,g') = \int_{\frac12+i\R} d\Delta \int_\C d^{(2)}z\ \Phi^\Delta_{z,\bz}(g)\Phi^{\Delta^*}_{z,\bz}(g')\ ,
\label{dgg}
\eea
where the reflected dimension $\Delta^*=1-\Delta$ was defined in eq. (\ref{dod}). The equivalence between representations with labels $\Delta$ and $\Delta^*$ manifests itself in the relation $C_2(\Delta^*)=C_2(\Delta)$, and in the reflection relation
\bea
\Phi^\Delta_{z,\bz}(g) = R(\Delta)\int_\C d^{(2)}z'\ \left|{\cal E}(\Delta,\Delta|z,z')\right|^2 \Phi^{\Delta^*}_{z',\bz'}(g)\ ,
\label{psp}
\eea
where $R(\Delta)$ is a reflection coefficient, and the two-point invariant ${\cal E}(\Delta,\Delta|z,z')$ was given in eq. (\ref{edd}).

Let us go back to Liouville theory, of which $SL_2(\C)$ quantum mechanics is the large $c$ limit. 
An important axiom of conformal field theory is the assumption that a four-point function can be decomposed into four-point conformal blocks,
\bea
\la \prod_{i=1}^4 V_{\Delta_i}(z_i,\bz_i) \ra  = \int d\Delta_s\ C(c|\Delta_1,\Delta_2,\Delta_s)C(c|\Delta_s,\Delta_3,\Delta_4) \left|{\cal G}_{\Delta_s}(c|\Delta_i|z_i)\right|^2 \ ,
\label{ccg}
\eea
where the structure constant $C(c|\Delta_1,\Delta_2,\Delta_3)$ depends on the central charge $c$ and on the conformal dimensions $\Delta_i$, but not on the field positions $z_i$. (We will not study how the $c$-dependent contour of integration must be manipulated in the large $c$ limit.) We now check that our formula (\ref{fds}) for the large $c$ four-point block ${\cal F}_{\Delta_s}(\Delta_i|z_i)$ is compatible with the large $c$ limit of this axiom. To do this, we insert the identity $1=\int_{SL_2(\C)} dg'\ \delta(g,g')$ together with the formula (\ref{dgg}) for $\delta(g,g')$ in the large $c$ limit (\ref{oip}) of the four-point function, 
\begin{multline}
\underset{c\rar \infty}{\lim} \la \prod_{i=1}^4 V_{\Delta_i}(z_i,\bz_i)\ra = \int_{\frac12+i\R} d\Delta_s \int_\C d^{(2)}z_s\ 
\\
\int_{SL_2(\C)} dg\ \left(\Phi^{\Delta_1}_{z_1,\bz_1}\Phi^{\Delta_2}_{z_2,\bz_2}\Phi^{\Delta_s}_{z_s,\bz_s} \right)(g)
\int_{SL_2(\C)} dg'\ 
\left(\Phi^{\Delta^*_s}_{z_s,\bz_s}\Phi^{\Delta_3}_{z_3,\bz_3}\Phi^{\Delta_4}_{z_4,\bz_4} \right)(g') \ .
\end{multline}
The integrals over $g$ and $g'$ produce $z$-dependent factors proportional to the three-point invariant ${\cal E}(\Delta_1,\Delta_2,\Delta_3|z_1,z_2,z_3)$ (\ref{eddd}), and $z$-independent factors which we call $B(\Delta_1,\Delta_2,\Delta_3)$,
\begin{multline}
\underset{c\rar \infty}{\lim} \la \prod_{i=1}^4 V_{\Delta_i}(z_i,\bz_i)\ra = \int_{\frac12+i\R} d\Delta_s\ B(\Delta_1,\Delta_2,\Delta_s)B(\Delta_s^*,\Delta_3,\Delta_4)
\\
\times \int_\C d^{(2)}z_s\ \left|{\cal E}(\Delta_1,\Delta_2,\Delta_s|z_1,z_2,z_s){\cal E}(\Delta_s^*,\Delta_3,\Delta_4|z_s,z_3,z_4)\right|^2\ .
\end{multline}
Decomposing the integral $\int_\C d^{(2)}z_s$ into a combination of contour integrals over $z_s$ and $\bz_s$ yields a linear combination of the two terms $\left|{\cal F}_{\Delta_s}(\Delta_i|z_i)\right|^2$ and $\left|{\cal F}_{\Delta_s^*}(\Delta_i|z_i)\right|^2$ where ${\cal F}_{\Delta_s}(\Delta_i|z_i)$ is given by eq. (\ref{fds}). But these two terms give the same contribution to $\underset{c\rar \infty}{\lim} \la \prod_{i=1}^4 V_{\Delta_i}(z_i,\bz_i)\ra$, because the integration contour in $\int_{\frac12+i\R} d\Delta_s$ is invariant under the reflection $\Delta_s\rar \Delta_s^*$. Absorbing any remaining prefactors into the $B$-factors, we obtain
\bea
\underset{c\rar \infty}{\lim} \la \prod_{i=1}^4 V_{\Delta_i}(z_i,\bz_i)\ra = \int_{\frac12+i\R} d\Delta_s\ B(\Delta_1,\Delta_2,\Delta_s)B(\Delta_s^*,\Delta_3,\Delta_4) \left|{\cal F}_{\Delta_s}(\Delta_i|z_i)\right|^2\ .
\eea
This formula can be interpreted as the large $c$ limit of the decomposition (\ref{ccg}) of a four-point function into four-point conformal blocks, provided we have $B(\Delta_1,\Delta_2,\Delta_3)=\underset{c\rar \infty}{\lim} C(c|\Delta_1,\Delta_2,\Delta_3)$.
This provide a justification for the formula (\ref{fds}) for ${\cal F}_{\Delta_s}(\Delta_i|z_i)$.

\subsection{Point particle on $SL_3(\C)$ \label{sapp}}

A basis of functions on $SL_3(\C)$ can be defined as 
\bea
\Phi^j_{Z,\bar{Z}}(g) = \left[u_Z P g^{-1T}P u_{\bar{Z}}^T\right]^{-r} \left[v_Z g v_{\bar{Z}}^T\right]^{-s},\quad \left\{\begin{array}{l} u_Z = (w,-x,1) \\ v_Z = (xy-w,-y,1) \end{array}\right.,\quad P=\bsm 0 & 0 & 1 \\ 0 & -1 & 0 \\ 1 & 0 & 0 \esm \ ,
\label{pzg}
\eea
where we recall that $Z=(w,x,y)$ is a three-component isospin vector, and that the components $(r,s)$ of the spin $j$ are defined in eq. (\ref{rjsj}). The vectors $u_Z$ and $v_Z$ are such that
\bea
u_Z \om(t^a)  = D_{(-h_1,Z)}(t^a) u_Z \scs v_Z t^a = D_{(h_3,Z)}(t^a) v_Z  \ ,
\eea
where the differential operators $D_{(j,Z)}(t^a)$ are defined in eqs. (\ref{dh})-(\ref{df}) and the weights $h_i$ in eq. (\ref{hhh}), 
the $s\ell_3$ generators $t^a$ are represented as the matrices
\bea
&& h^1=\bsm 1 &0 &0 \\ 0& -1 & 0\\ 0&0 & 0 \esm,\ h^2 = \bsm 0 &0 &0 \\ 0& 1 & 0\\ 0& 0& -1 \esm,
\label{slh}
\\
&& e^1=\bsm 0& 1 & 0\\ 0&0 & 0 \\ 0& 0& 0\esm,\ e^2 = \bsm 0& 0 &0 \\ 0& 0& 1 \\ 0& 0&0 \esm,\ e^3 = \bsm 0& 0& 1 \\ 0& 0&0\\ 0&0 & 0\esm,\
\\
&&
f^1=\bsm 0 &0 &0 \\ 1 &0 & 0\\ 0& 0 &0 \esm,\ f^2 = \bsm 0& 0&0 \\ 0 &0 & 0\\ 0& 1 &0 \esm,\ f^3 = \bsm 0&0 &0 \\ 0& 0&0 \\ 1 & 0&0 \esm\ ,
\label{slf}
\eea
and the action of the Dynkin diagram automorphism $\om$ on such matrices is
\bea
\om(t^a) = -P(t^a)^T P \ .
\eea
It follows that the function $\Phi^j_{Z,\bar{Z}}(g)$ behaves under $s\ell_3$ tranformation as
\bea
\Phi^j_{Z,\bar{Z}}((1+\e t^a)g) = (1 + \e D_{(j,Z)}(t^a)) \Phi^j_{Z,\bar{Z}}(g) + O(\e^2) \ .
\eea
As a result, $\Phi^j_{Z,\bar{Z}}(g)$ is an eigenvector of the quadratic (Laplacian) and cubic invariant differential operators on $SL_3(\C)$, with the respective eigenvalues
\bea
C_2(j) &=&  (j,j+2e_1+2e_2) =  \tfrac23(r^2+s^2+rs)-2r-2s \ , 
\label{kk}
\\ 
C_3(j) &=&  -6\prod_{i=1}^3 (h_i,j+e_1+e_2) =\tfrac29 (r-s)(2r+s-3)(2s+r-3) \ .
\label{kkk}
\eea
These Casimir numbers can be derived by computing $s\ell_3$-invariant combinations of the differential operators $D_{(j,Z)}(t^a)$, using the covariant tensors 
\bea
g^{ab} = \Tr t^at^b \scs d^{abc} = \Tr (t^a t^b t^c + t^a t^c t^b) \ .
\eea
Then we have the identities
\bea
g_{ab}D_{(j,Z)}(t^a)D_{(j,Z)}(t^b) &=& C_2(j) \ ,
\label{gi}
\\ 
d_{abc}D_{(j,Z)}(t^a)D_{(j,Z)}(t^b)D_{(j,Z)}(t^c) &=& C_3(j)\  .
\label{di}
\eea
The Casimir numbers $C_2(j)$ and $C_3(j)$ are invariant under the six Weyl reflections (\ref{rsr}), which is a manifestation of the equivalence of two representations whenever their spins are related by a reflection. At the level of the function $\Phi^j_{Z,\bar{Z}}(g)$, this equivalence manifests itself as a relation which we now write in the case of the maximal reflection $j\rar j^*$,
\bea
\Phi^j_{Z,\bar{Z}}(g) = R(j) \int dZ'd\bar{Z}'\ \left|{\cal E}(j,j^\om|Z,Z')\right|^2 \Phi^{j^*}_{Z',\bar{Z}'}(g) \ ,
\label{prp}
\eea
where $R(j)$ is a reflection coefficient, and the two-point invariant ${\cal E}(j,j^\om|Z,Z')$
was given in eq. (\ref{fj}). This generalizes the $SL_2(\C)$ reflection relation eq. (\ref{psp}).

Now, the function $\Phi^j_{Z,\bar{Z}}(g)$ simplifies if either $r=0$ or $s=0$, and then it obeys the differential equations,
\bea
d^{(1)}_Z \Phi^{(0,s)}_{Z,\bar{Z}}(g) = d^{(1)}_{\bar{Z}} \Phi^{(0,s)}_{Z,\bar{Z}}(g) = 0 \scs d^{(2)}_Z \Phi^{(r,0)}_{Z,\bar{Z}}(g) = d^{(2)}_{\bar{Z}} \Phi^{(r,0)}_{Z,\bar{Z}}(g)  = 0 \ ,
\label{dpd}
\eea
where the operators $d^{(1)}_Z$ and $d^{(2)}_Z$ were defined in eq. (\ref{dzdz}). This is a consequence of the vectors $u_Z$ and $v_Z$ obeying $d^{(2)}_Z u_Z = d^{(1)}_Z v_Z=0$, and this justifies our definitions of $d^{(1)}_Z$ and $d^{(2)}_Z$. Notice that these operators 
obey the remarkable property
\bea
d^{(k)}_Z D_{(0,Z)}(t^a) = D_{(-e_k,Z)}(t^a) d^{(k)}_Z\ ,  \qquad (k=1,2) \ .
\eea
(We recall that $e_k$ are the simple roots of $s\ell_3$; for instance the coordinates of the spin $j=-e_1$ are $(r,s)=(2,-1)$.)

Finally, we can write the analog in $s\ell_3$ conformal Toda theory of the expression (\ref{oip}) for large $c$ Liouville theory correlation functions, by using the functions $\Phi^j_{Z,\bar{Z}}(g)$ on $SL_3(\C)$. The isospin variable $Z$ must be specialized as $Z=\bar{z}=(\frac12 z^2,z,z)$ (as in eq. (\ref{zz})), and we must remember the relation (\ref{baj}) between $W_3$ momenta $\al$ and $s\ell_3$ spins $j$. Calling $V_\al(z,\bz)$ the vertex operator of $s\ell_3$ conformal Toda theory with the momentum $\al$, we have \cite{fl07c}
\bea
\underset{c\rar \infty}{\lim}
\la \prod_{i=1}^n V_{\al_i}(z_i,\bz_i) \ra = \int_{SL_3(\C)}dg\ \prod_{i=1}^n \Phi^{j_i}_{\vec{z}_i,\vec{\bz}_i}(g) \ .
\label{lvp}
\eea

\zeq\section{Derivation of the expansion of a four-point block \label{appder}}

Here we derive the expansion eq. (\ref{fsss}) of the large $c$ four-point block ${\cal F}_{j_s}(z)={\cal F}_{j_s}(j_i|0,z,1,\infty)$, starting from the integral formula eq. (\ref{fgg}). We propose two possible ways to perform the calculation. The first way is more straightforward, but it leads to a formula (\ref{fres}) which is less symmetric than eq. (\ref{fsss}) and has spurious poles. The second way is less straightforward as it starts with a six-dimensional (instead of three-dimensional) integral, but the symmetry $(\al,\delta,s)\lrar (\beta,\g,r)$ of eq. (\ref{fsss}) is manifest throughout the calculation. (We recall that $r$ and $s$ are the components of $j_s$, and that the combinations of spin components $\al,\beta,\g,\delta$ are defined in eq. (\ref{abcd}).)

\subsection{First way}


After performing a few change of variables, the integral formula (\ref{fgg}) (together with eq. (\ref{ez})) leads to 
\begin{multline}
{\cal F}_{j_s}(z) =  z^{-\Delta_{12}^s} e^{i\pi (\beta-\delta+s)}
\frac{\G(r)\G(s)\G(r+s-1)}{\G(\beta+s-1)\G(\delta)\G(-\delta+r+s-1)\G(r-\beta)}  
\int_{C_0} dwdxdy\
\\ w^{\delta-s}(xy-w)^{\beta+s-2}(y+w-xy)^{-\beta}(w-x+1)^{-\delta+r+s-2}(1-yz)^{-\al}(1-xz+wz^2)^{-\gamma} \ ,
\label{fgi}
\end{multline}
where the condition (\ref{fzo}) has been used for determining the prefactor and the integration contour 
\bea
C_0\ : \quad y\in(\tfrac{w}{x},\tfrac{w}{x-1}) \quad {\rm then} \quad w\in(x-1,0) \quad {\rm then}\quad  x\in(0,1)\ .
\eea
Let us denote the integral (\ref{fgi}) as $ {\cal F}_{j_s}(z)= \la (1-yz)^{-\al}(1-xz+wz^2)^{-\gamma}\ra$, and expand the integrand in powers of $z$. This reduces the problem to computing expectation values of monomials $w^n x^k y^m$, and we find
\bea
\la w^n x^k y^m\ra = z^{-\Delta_{12}^s} (-1)^n \frac{(\delta)_{m+n}(r-\beta)_n}{(r+s-1)_{m+n}} \sum_{\ell=0}^m C^\ell_m \frac{(\beta-\ell)_k}{(s)_\ell(r)_{k+n-\ell}}\ ,
\eea
where the notation $(t)_n$ was introduced in eq. (\ref{tn}), and we write $C^\ell_m=\frac{m!}{\ell!(m-\ell)!}$.
This leads to 
\bea
{\cal F}_{j_s}(z) =  z^{-\Delta_{12}^s} \sum_{q,m=0}^\infty \frac{(\g)_q(\al)_m}{q!m!} \sum_{i=0}^q\sum_{\ell=0}^m z^{q+m+i}C^i_qC^\ell_m \frac{(r-\beta)_i(\beta-\ell)_{q-i}(\delta)_{m+i}}{(s)_\ell(r)_{q-\ell}(r+s-1)_{m+i}}\ .
\label{fres}
\eea
This can be seen to agree with eq. (\ref{fsss}) by an automatic calculation of the first few orders in $z$. This agreement is non-trivial, as it involves the cancellation of the spurious poles which are present in eq. (\ref{fres}).

\subsection{Second way}

We start again with eq. (\ref{fgg}). We fist want to make this formula more symmetric, at the expense of replacing the three-dimensional integral over $Z_s$ with a six-dimensional integral. This is done by using a reflection relation for conformal blocks, which is the holomorphic half of the reflection relation eq. (\ref{prp}) for functions on $SL_3(\C)$. This leads to 
\bea
{\cal F}_{j_s}(j_i|z_i) &=& {\cal N}_1\int_{C_1} dZ_sdZ'_s\ {\cal E}(j_1,j_2,j_s|\vec{z}_1,\vec{z}_2,Z_s) {\cal E}(j_s^{*\om},j_s^*|Z_s,Z'_s) {\cal E}(j_s^{\om},j_3,j_4|Z'_s,\vec{z}_3,\vec{z}_4)\ ,
\\
&=& {\cal N}_1 
\rho_{34}^{s-\al-r_4}\rho_{12}^{r-\beta-r_1} \int_{C_1} dZ_s dZ'_s\ \chi_{34s}^{-\al}\rho_{s4}^{\al-s}\rho_{4s}^{\g-r}\rho_{3s}^{-\g}\ \rho_{ss'}^{s-2}\rho_{s's}^{r-2}\ \chi_{12s'}^{-\beta}\rho_{s'1}^{\beta-r}\rho_{1s'}^{\delta-s}\rho_{2s'}^{-\delta}\ .
\label{nizz}
\eea
The notations come from Subsection \ref{ssif}, except the definitions of $\vec{z}$ (\ref{zz}) and of $\al,\beta,\g,\delta$ (\ref{abcd}). The contour $C_1$ and normalization factor ${\cal N}_1$ are supposed to be determined by the condition (\ref{fzo}). We will not keep track of contours and normalizations explicitly; instead we will call $C_i$ and ${\cal N}_i$ the various contours and normalizations which appear in the calculation. 

Out of the six integrals $\int dZ_sdZ'_s$ in eq. (\ref{nizz}), the two integrals over $y_s$ and $y'_s$ can be performed immediately using the formula $\int_C dy\ \prod_{i=1}^3 (a_iy-b_i)^{\al_i} = {\cal N} \prod_{i<j} (a_ib_j-a_jb_i)^{\al_i+\al_j+1}$ (assuming $\sum_{i=1}^3 \al_i=-2$), where the choice of the integration contour $C$ only affects the normalization factor ${\cal N}$ and is therefore not important for us. Then, we replace the four remaining variables $(w_s,x_s,w'_s,x'_s)$ with four new variables $(w,x,w',x')$, using the change of variables
\bea
x_s &=& \frac{\frac{z_4z_{12}}{z_{24}}x + \frac{z_1z_{34}}{z_{31}} -\frac12 (z_1+z_4)z(w+x+1)}{\frac{z_{12}}{z_{42}} x +\frac{z_{34}}{z_{31}} - z(w+x+1)} \ ,
\\
w_s &=& \frac12\frac{\frac{z_4^2z_{12}}{z_{42}}x +  \frac{z^2_1z_{34}}{z_{31}} - z_1z_4z(w+x+1)}{ \frac{z_{12}}{z_{42}} x +\frac{z_{34}}{z_{31}} - z(w+x+1)}\ ,
\eea
and similarly for $x'_s$ and $z'_s$, with the exchanges of indices $\bsm 1\lrar 4 \\ 2\lrar 3 \esm$.
We can then check global conformal invariance, and restrict our attention to ${\cal F}_{j_s}(z) = {\cal F}_{j_s}(j_i|0,z,1,\infty)$ as in eq. (\ref{fog}), 
\begin{multline}
{\cal F}_{j_s}(z) = z^{-\Delta_{12}^s} {\cal N}_2\int_{C_2} dxdwdx'dw'\ (w'-wx'+1-xx')^{\al-1}(w-w'x+1-xx')^{\beta-1}
\\ \times (w'-wx')^{-\al+s-1} (w-w'x)^{-\beta+r-1} (1-z(w+1))^{-\g}(1-z(w'+1))^{-\delta}\ .
\end{multline}
We now perform another change of integration variables, introducing new variables $\sigma,\tau$ which this time mix the unprimed $(x,w)$ and primed $(x',w')$ variables:
\bea
\sigma = w'-wx' \scs \tau = w-w'x \ .
\eea
We also introduce the notation $\om = 1-xx'$ for convenience, and we obtain
\begin{multline}
{\cal F}_{j_s}(z) = z^{-\Delta_{12}^s} {\cal N}_3\int_{C_3} dxdx'd\sigma d\tau\ \om^{-1}(\sigma+\om)^{\al-1}(\tau+\om)^{\beta-1}\sigma^{-\al+s-1}\tau^{-\beta+r-1}
\\
\times \left(1-\tfrac{z}{\om}(x\sigma+(\tau+\om))\right)^{-\g} \left(1-\tfrac{z}{\om}(x'\tau+(\sigma+\om))\right)^{-\delta}\ .
\end{multline}
We expand the last two factors, for instance
\bea
\left(1-\tfrac{z}{\om}(x\sigma+(\tau+\om))\right)^{-\g} = \sum_{n,i=0}^\infty
\frac{(\g)_{n+i}}{n!i!} \left(\tfrac{z}{\om}\right)^{n+i} (x\sigma)^n (\tau+\om)^i \ ,
\eea
where the notation $(t)_n$ was introduced in eq. (\ref{tn}). Then we integrate over $\sigma,\tau\in (0,\infty)$, and then over $x,x'\in \C$ such that $\bar{x}'=-x$ using $\int_{\bar{x}'=-x}dxdx'\ (1-xx')^{t-2} x^nx'^{n'} = {\cal N} \delta_{n,n'}\frac{n!}{(t)_n}$, where ${\cal N}$ is an $n$-independent normalization factor. This directly leads to eq. (\ref{fsss}).


\begin{thebibliography}{10}
\expandafter\ifx\csname url\endcsname\relax
  \def\url#1{{\tt #1}}\fi
\expandafter\ifx\csname urlprefix\endcsname\relax\def\urlprefix{URL }\fi
\providecommand{\eprint}[2][]{\url{#2}}

\bibitem{bpz84}
A.~A. Belavin, A.~M. Polyakov, A.~B. Zamolodchikov, {\em Infinite conformal
  symmetry in two-dimensional quantum field theory\/}, Nucl. Phys. B241 pp.
  333--380 (1984)

\bibitem{aflt10}
V.~A. Alba, V.~A. Fateev, A.~V. Litvinov, G.~M. Tarnopolsky, {\em {On
  combinatorial expansion of the conformal blocks arising from AGT
  conjecture}\/}, Lett. Math. Phys. 98 pp. 33--64 (2011), \eprint{1012.1312}

\bibitem{agt09}
L.~F. Alday, D.~Gaiotto, Y.~Tachikawa, {\em {Liouville Correlation Functions
  from Four-dimensional Gauge Theories}\/}, Lett. Math. Phys. 91 pp. 167--197
  (2010), \eprint{0906.3219}

\bibitem{zam85}
A.~B. Zamolodchikov, {\em {Infinite Additional Symmetries in Two-Dimensional
  Conformal Quantum Field Theory}\/}, Theor. Math. Phys. 65 pp. 1205--1213
  (1985)

\bibitem{fz86b}
V.~A. Fateev, A.~B. Zamolodchikov, {\em Conformal quantum field theory models
  in two dimensions having $Z_3$ symmetry\/}, Nucl. Phys. B280 pp. 644--660
  (1987)

\bibitem{fl88}
V.~A. Fateev, S.~L. Lukyanov, {\em The Models of Two-Dimensional Conformal
  Quantum Field Theory with Z(n) Symmetry\/}, Int. J. Mod. Phys. A3 p. 507
  (1988)

\bibitem{fr10}
V.~Fateev, S.~Ribault, {\em {Conformal Toda theory with a boundary}\/}, JHEP 12
  p. 089 (2010), \eprint{1007.1293}

\bibitem{fl07c}
V.~A. Fateev, A.~V. Litvinov, {\em Correlation functions in conformal Toda
  field theory I\/}, JHEP 11 p. 002 (2007), \eprint{arXiv:0709.3806 [hep-th]}

\bibitem{fl08}
V.~A. Fateev, A.~V. Litvinov, {\em {Correlation functions in conformal Toda
  field theory II}\/}, JHEP 01 p. 033 (2009), \eprint{0810.3020}

\bibitem{zam84}
Al.~Zamolodchikov, {\em {Conformal symmetry in two dimensions: an explicit
  recurrence formula for the conformal partial wave amplitude}\/},
  Commun.Math.Phys. 96 pp. 419--422 (1984)

\bibitem{fl11}
V.~Fateev, A.~Litvinov, {\em {Integrable structure, W-symmetry and AGT
  relation}\/}  (2011), * Temporary entry *, \eprint{1109.4042}

\bibitem{fms97}
P.~Di~Francesco, P.~Mathieu, D.~Senechal, {\em Conformal field theory\/} New
  York, USA: Springer (1997) 890 p

\bibitem{fgg75}
S.~Ferrara, R.~Gatto, A.~Grillo, {\em {Properties of Partial Wave Amplitudes in
  Conformal Invariant Field Theories}\/}, Nuovo Cim. A26 p. 226 (1975)

\bibitem{do93}
F.~Dolan, H.~Osborn, {\em {Conformal partial waves and the operator product
  expansion}\/}, Nucl.Phys. B678 pp. 491--507 (2004), \eprint{hep-th/0309180}

\bibitem{bw91}
P.~Bowcock, G.~M.~T. Watts, {\em On the classification of quantum W
  algebras\/}, Nucl. Phys. B379 pp. 63--95 (1992), \eprint{hep-th/9111062}

\end{thebibliography}

\end{document}